\documentclass[
 reprint,
bibnotes,
 amsmath,amssymb,
 aps,
floatfix,dvipdfmx
]{revtex4-1}

\usepackage{tikz,pgfplots}
\usepackage{graphicx}
\usepackage{dcolumn}
\usepackage{bm}
\usepackage{braket}
\usepackage{cases}
\usepackage{amssymb}

\usepackage[colorlinks=true,linkcolor=black, allcolors=black,citecolor=black,urlcolor  = black,anchorcolor = black]{hyperref}%

\pgfplotsset{compat=1.14}

\begin{document}

\title{Scaling Hypothesis of Spatial Search on Fractal Lattice Using Quantum Walk}

\author{Rei Sato}%
\email{sato.r.av@m.titech.ac.jp}
\author{Tetsuro Nikuni}
\author{Shohei Watabe}
\affiliation{%
 Department of Physics, Faculty of Science Division I, Tokyo University of Science, Shinjuku, Tokyo,162-8601, Japan
}%
\begin{abstract} 
We investigate a quantum spatial search problem on fractal lattices, such as Sierpinski carpets and Menger sponges. 
In earlier numerical studies of the Sierpinski gasket, the Sierpinski tetrahedron, and the Sierpinski carpet, 
conjectures have been proposed for the scaling of  a quantum spatial search problem finding a specific target, which is given in terms of the characteristic quantities of a fractal geometry. 
We find that our simulation results for extended Sierpinski carpets and Menger sponges support the conjecture for the {\it optimal} number of the oracle calls, where the exponent is given by $1/2$ for $d_{\rm s} > 2$ and the inverse of the spectral dimension $d_{\rm s}$ for $d_{\rm s} < 2$. 
We also propose a scaling hypothesis for the {\it effective} number of  the oracle calls defined by the ratio of the {\it optimal} number of oracle calls to a square root of the maximum finding probability. 
The form of the scaling hypothesis for extended Sierpinski carpets is very similar but slightly different from the earlier conjecture for the Sierpinski gasket, the Sierpinski tetrahedron, and the conventional Sierpinski carpet. 
\end{abstract}
\maketitle


\section{\label{sec:Intro}Introduction}

The complex network is ubiquitous in many real-world systems, including the World-Wide-Web, social and biological networks, such as the actors network, protein-protein interaction network, and cellular network~\cite{song2005self}. 
Contrary to conventional wisdom that the complex networks are not the self-similar due to the ``small-world" property, the renormalization technique has revealed that the structure of the complex network is indeed self-similar~\cite{song2005self}. 
The efficient database search on such networks is an important issue, and one of the efficient schemes is to apply the spatial search using quantum walk, which has been applied to a random network, the so-called Erd\"os--Ranyi random graphs~\cite{chakraborty2016spatial}. 
In this respect, studying the quantum spatial search on fractal geometries is relevant to practical application to real-world systems composed of complex networks with self-similarity, and yet there naturally arises a fundamental question; what kind of scaling law is behind the quantum spatial search on self-similar networks. The fractal lattice, such as the Sierpinski gasket,  Sierpinski carpet, or Menger sponge, is one of the simplest but most fundamental structures to investigate this problem.

A spatial search problem is a kind of database search to find a marked item from a given $N$-site graph. 
A quantum search algorithm \cite{grover1997quantum} using quantum walk provides faster performance than classical search algorithm \cite{Shenvi:2003beb}. 
This so-called Grover's algorithm, as well as the Shor's algorithm, are well-known algorithms based on quantum mechanics~\cite{Shor1994}. 
Nowadays the quantum walk can be experimentally implemented in various systems, such as an optical lattice~\cite{Karski:2009jc}, an ion trap~\cite{Zahringer:2010bs}, and photonic systems~\cite{Xiao2018it,Xiao2017iw,Schreiber2010cl,Schreiber2012wj}.

The idea of the Grover's algorithm is to amplify the probability finding a target point~\cite{grover1997quantum}. It is achieved by repeatedly applying a quantum oracle operator and then applying the Grover's diffusion operator. 
The quantum oracle operator changes the sign of the probability amplitude of the target site, and the Grover's diffusion operator provides the inversion about the mean. Since it obeys a unitary evolution, the dynamics of the probability finding a specific target is periodic as a function of the number of oracle calls. 
A helpful intuitive picture for understanding the quantum spatial search is a Schr\"odinger dynamics on a lattice with a single attractive potential at the target site~\cite{Grover2001}.

After the Grover's paper~\cite{grover1997quantum}, the quantum search has been intensively and extensively studied~\cite{Shenvi:2003beb,Grover2001,childs2004spatial,
aaronson2003quantum,Childs2004Dirac,ambainis2005coins,Patel2010I,tulsi2008faster,Agliari2010dw,Patel2010II,Lovett2011cv,patel2012search,Foulger2015,Boettcher2018_012309,Boettcher2018_012320,Osada2018ej,tamegai2018spatial,Abhijith2018}. 
One of the main issues is to understand the scaling relation between the optimal number of oracle calls $Q$ and the number of sites $N$.  
The scaling behavior varies strongly depending on the spatial dimension of the graph. 
For $d=1$, the scaling law is $Q=\mathcal{O}(N)$, which means that the efficiency is equivalent to that of classical search~\cite{childs2004spatial,Patel2010I}.  
For $d> 2$, the scaling law is $Q=\mathcal{O}(\sqrt{N})$ in a hypercubic lattice for the continuous-time or discrete-time algorithm~\cite{aaronson2003quantum,Childs2004Dirac,ambainis2005coins,Patel2010I}, which is consistent with the original Grover's search~\cite{grover1997quantum}. 
On the other hand, the two-dimensional system is critical, which gives the scaling law in the form $Q = \mathcal{O}(\sqrt{N}\ln^{\epsilon}{N})$~\cite{aaronson2003quantum,ambainis2005coins,Childs2004Dirac,childs2004spatial,tulsi2008faster,Patel2010I}. 
Indeed, for $d=2$, the relations $Q=\mathcal{O}(\sqrt{N}\ln^{3/2}{N})$~\cite{aaronson2003quantum} and $\mathcal{O}(\sqrt{N}\ln^{}{N})$~\cite{ambainis2005coins,Childs2004Dirac} are reported. 
In $d =2$, the faster search algorithm, which gives $Q={\mathcal O}(\sqrt{N\ln{N}})$, is also proposed by Tulsi, which employs an ancilla qubit~\cite{tulsi2008faster,Patel2010I}. 

From these results, Patel and Raghunathan argued that the scaling law for spatial search in $d$-dimension obeys~\cite{Patel2010I}
\begin{equation}
\label{eq:scaling formula 1}
    Q \ge {\rm max}\{dN^{1/d},\pi\sqrt{N}/4\}, 
\end{equation} 
except for $d=2$, where emerges the logarithmic slowing down analogous to critical phenomena in statistical mechanics~\cite{Patel2010II}. 
Based on this formula, they posed the following questions:  does the relation (\ref{eq:scaling formula 1}) hold in the non-integer dimensional system? 
If so, what is an appropriate dimension $d$ appearing in (\ref{eq:scaling formula 1})?  In general, the fractal system is characterized by following three kinds of the dimension: the Euclidean dimension, the fractal dimension, and the spectral dimension; definitions of these dimensions will be given in the next section.

Using the Sierpinski gasket and Sierpinski tetrahedron, Patel and Raghunathan numerically found that $d$ is given by the spectral dimension $d_{\rm s}$, but not the fractal dimension; they proposed the conjecture of the scaling law~\cite{patel2012search}
\begin{equation}
    Q \ge {\rm max}
    \{N^{1/d_{\rm s}},\pi\sqrt{N}/4\}.
    \label{scaling formula 2}
\end{equation} 
In order to check the validity of this conjecture, Tamegai and two of the authors recently investigated the scaling law (\ref{scaling formula 2}) by using the Sierpinski carpet~\cite{tamegai2018spatial}, which has the fractal and spectral dimensions different from the Sierpinski gasket and Sierpinski tetrahedron. They found that the scaling law is consistently given by the spectral dimension rather than by the fractal dimension in the case of the Sierpinski carpet. 

Furthermore, they proposed a novel scaling hypothesis about the effective number of the oracle calls, defined by $Q/\sqrt{P_{\rm max}}$ in the fractal lattice, where $P_{\rm max}$ is the maximum probability at the marked site. They found that the scaling exponent $\gamma$, given by $Q/\sqrt{P_{\rm max}} = {\mathcal O}(N^\gamma)$, is expressed by a combination of characteristic quantities of a fractal structure~\cite{tamegai2018spatial}: 
\begin{equation}
\gamma = \frac{d_{\rm s}}{d_{\rm E} -1} + d_{\rm f} -s, 
    \label{scaling formula 3}
\end{equation} 
which involves the Euclidean dimension $d_{\rm E}$, the fractal dimension $d_{\rm f}$, the spectral dimension $d_{\rm s}$, and the scaling factor $s$. 
This relation holds very well within the margin of error in the Sierpinski carpet~\cite{tamegai2018spatial}.  Furthermore, the same relation was shown to hold excellently also  
in the Sierpinski gasket and Sierpinski tetrahedron by using the numerical data reported by Patel and Raghunathan~\cite{patel2012search}. 

In this study, based on the studies of Refs.~\cite{patel2012search,tamegai2018spatial}, we further investigate the scaling hypotheses proposed in Refs.~\cite{patel2012search,tamegai2018spatial} in various kinds of fractal lattices: extended Sierpinski carpets, a conventional Menger sponge, and extended Menger sponges. 
In the extended Sierpinski carpets, the scaling hypothesis for the optimal number of oracle calls (\ref{scaling formula 2}) holds very well. 
In the Menger sponge case, where the Euclidean and fractal dimensions are greater than $2$, i.e., $d_{\rm E,f} > 2$, the scaling exponent gets saturates at the value close to $1/2$. 
In the extended Sierpinksi carpets, although the scaling hypothesis (\ref{scaling formula 3}) does not exactly hold, we find a variant of the scaling law very similar to (\ref{scaling formula 3}), which fits our numerical data very well. 

This paper is organized as follows. 
Section~\ref{sec:fractal} introduces quantities characterizing a fractal geometry, such as the Euclidean dimension, fractal dimension, spectral dimension, and scaling factor. 
Section~\ref{sec:quantum search algorithm} describes the spatial search algorithm with the discrete-time quantum walk. 
Section~\ref{sec:results} provides numerical simulation results, and discuss the scaling hypothesis. 
Finally, we conclude our results in Sec.~\ref{sec:conclusion}.

\section{\label{sec:fractal}Fractal geometry}

\begin{figure}[tbp]
      \centering
      \includegraphics[width=60mm]{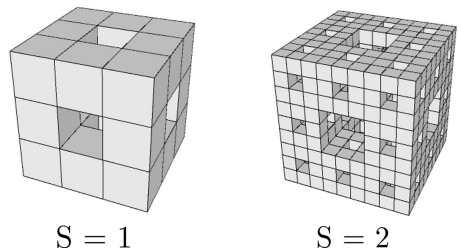}
      \caption{Menger sponges ${\rm MS}(3,1)$ at stages $S = 1$ and $2$, where the scaling factor is $s=3$. 
      }\label{menger sponge}
\end{figure}

\begin{figure}[tb]
      \centering
      \includegraphics[width=90mm]{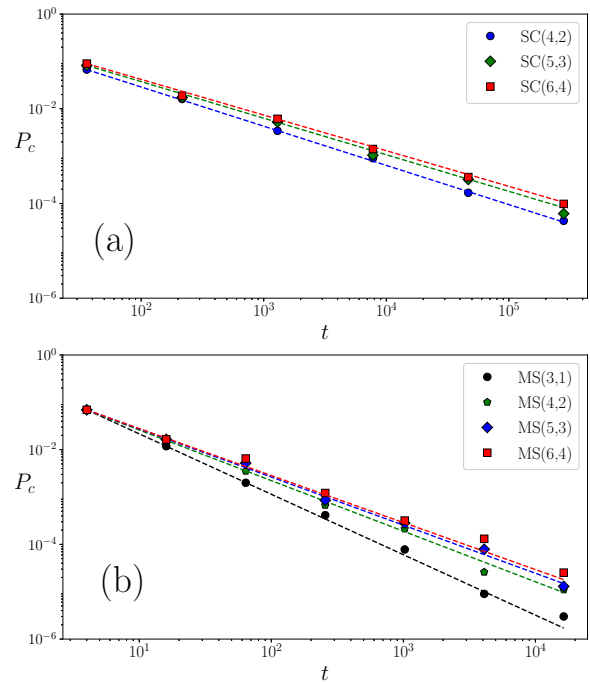}
      \caption{(Color online) Return probability of a classical random walker on fractal lattices: (a) Sierpinski carpets and (b) Menger sponges. 
      (a) For Sierpinski carpets, we used a fixed boundary condition, and evaluated the spectral dimension $d_{\rm s}$ from data at the stage $S = 8$ for ${\rm SC}(4,2)$, at $S=7$ for ${\rm SC}(5,3)$, and at $S=6$ for ${\rm SC}(6,4)$. 
      (b) For Menger sponges, we employed a periodic boundary condition, and evaluated the spectral dimension $d_{\rm s}$ from data 
      at $S=5$ for ${\rm MS}(3,1)$, and at $S=4$ for ${\rm MS}(4,2), {\rm MS}(5,3)$ and ${\rm MS}(6,4)$.}\label{random walk}
\end{figure}

\begin{table*}[htbp]
  \centering
    \renewcommand{\arraystretch}{1.2}
    {\tabcolsep=1.8mm
      \begin{tabular}{lccccccrrrrrrrr}\hline\hline
&${d_{\rm E}}$&${d_{\rm f}}$&${d_{\rm s}}$& $d_{\rm s}$\cite{barlow1999brownian} &${M(s)}$&${s}$&${s'}$&$L$&$N_{S=1}$&$N_{S=2}$&$N_{S=3}$&$N_{S=4}$&$N_{S=5}$\\ \hline
SC(4,2)& 2& 1.792& 1.62(2)&$1.56\cdots < d_{\rm s} < 1.67\cdots$& 12&4&2&${4^S}$&12& ${12^2}$& ${12^3}$& ${12^4}$ &${12^5}$\\  
SC(5,3)& 2& 1.723& 1.61(2)&$1.50\cdots < d_{\rm s} < 1.62\cdots$& 16 &5&3&${5^S}$&16& ${16^2}$& ${16^3}$& ${16^4}$&${16^5}$\\  
SC(6,4)&2&1.672&1.51(1)&$1.46\cdots < d_{\rm s} <  1.55\cdots$&20&6&4&${6^S}$&20& ${20^2}$&${20^3}$& ${20^4}$&\\\hline
MS(3,1)&3& 2.727& 2.55(6)&$2.21\cdots < d_{\rm s} <  2.60\cdots$& 20 & 3 & 1&${3^S}$&20& ${20^2}$& ${20^3}$& ${20^4}$& ${20^5}$&\\ 
MS(4,2)& 3& 2.500& 2.12(5)&$2.00\cdots < d_{\rm s} <  2.26\cdots$& 32 & 4&2&${4^S}$&32&${32^2}$& ${32^3}$&${32^4}$&\\ 
MS(5,3)& 3& 2.351& 2.02(4)&$1.89\cdots < d_{\rm s} <  2.07\cdots$& 44 & 5&3&${5^S}$&44& ${44^2}$& ${44^3}$&${44^4}$&\\
MS(6,4)& 3& 2.247& 1.98(8)&$1.82\cdots < d_{\rm s} <  1.95\cdots$& 56 &6&4&${6^S}$&56& ${56^2}$& ${56^3}$&&\\\hline\hline
      \end{tabular}}
    \caption{Characteristic structure of fractal lattices of Sierpinski carpets (SC) and Menger sponges (MS). 
    The Euclidean dimensions $d_{\rm E}$, the fractal dimension $d_{\rm f}$, and the spectral dimension $d_{\rm s}$.  $L$ is a length of a side of fractal lattices at a stage $S$, and $N_S$ is the number of sites in a fractal lattice at a stage $S$.  In the blank cell, we do not evaluate it due to the large computation cost. }
    \label{dimension and target sites}
\end{table*} 

\begin{figure}[tbp]
    \centering
    \includegraphics[width=70mm]{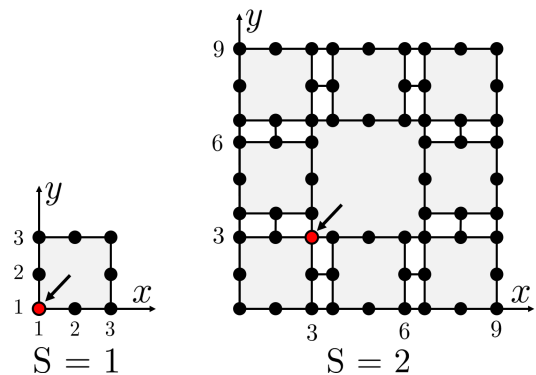}
    \caption{(Color online) Sierpinski carpet ${\rm SC}(3,1)$ at stages $S=1$ and $2$. A red vertex pointed by an arrow at each stage represents a target site ${\bf x}_0$ for the spatial search.}
    \label{SC(3,1)}
\end{figure}

\begin{figure}[tbp]
\begin{center}
          \includegraphics[width=45mm]{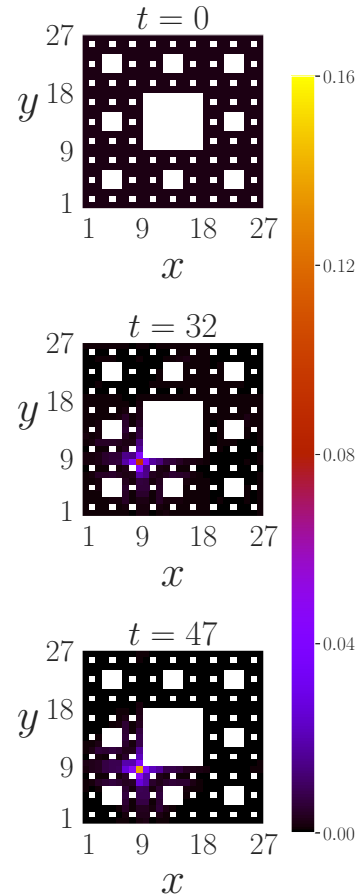}
    \caption{(Color online) Probability distribution of quantum spatial search on Sierpinski carpet SC$(3,1)$ at the stage ${S = 3}$.  The probability becomes highest at the marked vertex, when the time step reaches half the optimal number of oracle calls, $Q/2\simeq 47$. }
    \label{localized vertex}
              \end{center}
\end{figure}

Fractal geometry is ubiquitous in nature showing self-similar patterns~\cite{Mandelbrot:98509}. In this section, we briefly introduce quantities characterizing a fractal geometry, such as the Euclidean dimension, fractal dimension, spectral dimension as well as the scaling factor, which appear in scaling hypotheses discussed in this paper. 

The Euclid dimension $d_{\rm E}$ is a characteristic dimension of the Euclidean space, where the fractal geometry is embedded in. 
For the purpose of this study, it is sufficient to start with an equilateral triangle, a regular tetrahedron, a square, and a regular hexahedron. 
Suppose that $M(s)$ unit cells are generated in a structure by braking each side into $s$-pieces, where $s$ is called the scaling factor. 
If $M(s)=s^{d_E}$ is consistent for every scaling factor $s$, the scaling exponent $d_E$ is the Euclidean dimension, 
which gives $d_{\rm E} = \ln M(s) / \ln s$. 

For a self-similar pattern created by eliminating smallest unit pieces from the structure according to a certain  rule, 
the relation $M(s)=s^{d_E}$ no longer holds. 
In that case, one extends the definition of the dimension. If $M(s)=s^{d_{\rm f}}$ holds for arbitrary $s$ in the self-similar pattern, the scaling exponent, 
\begin{equation}
    d_{\rm f} = \frac{\ln M(s)}{\ln s}, 
\end{equation} 
is called the fractal dimension. 
In Menger sponge (Fig.~\ref{menger sponge}), for example, one has $d_{\rm f}=\ln{20}/\ln{3}=2.55\cdots$.

The spectral dimension $d_{\rm s}$ is related to a dynamical property of a classical random walk. 
One of the definitions is to relate $d_{\rm s}$ to a scaling behavior of the return probability through $P_{\rm c} ({\bf x}_0, t) \propto t^{-d_{\rm s}/2}$, where a random walker starts from a specific site ${\bf x}_0$ at $t=0$~\cite{Mayersbook,Klafterbook}. 
The Sierpinski gasket has an explicit form of the spectral dimension, given by $d_{\rm s}=2\ln{(d_E+1)}/\ln{(d_E+3)}$. 
However, since no analytic forms of the spectral dimension are known for Sierpinski carpets and Menger sponges at present, we numerically calculate the spectral dimensions from the scaling analysis of the return probability. 

For evaluating the spectral dimension, we consider the following discrete-time random walk. 
First, we randomly choose a single site ${\bf x}_0$ on the fractal lattice, where the walker starts the classical random walk.
At each time step, the walker moves $1$ unit in the randomly chosen direction (for example, in the case of Sierpinski carpet, we chose right, left, down, or up randomly with probability $1/4$ each).  If the nearest neighbor site does not exist in the chosen direction, the walker stays on the present site. 
Repeating this process, the diffusive behavior of the random walk on the fractal lattice can be analyzed. 
The return probability is estimated by the relation $P_{\rm c} ({\bf x}_0, t) = N ({\bf x}_0,t)/N_{\rm trial}$, 
where $N ({\bf x}_0,t)$ is the number of the events where the particle returns to the starting point ${\bf x}_0$ at a time $t$, and $N_{\rm trial}$ is the number of trials of the discrete-time random walk. 
Here, we take $N_{\rm trial} = 1,000,000$ in this study to evaluate the spectral dimension. 
We have determined the spectral dimension from the scaling fit as shown in Fig.~\ref{random walk}.

Table~\ref{dimension and target sites} summarizes characteristic quantities of fractal lattices discussed in the present paper. 
For Sierpinski carpets and Menger sponges, we use symbols ${\rm SC}(s,s')$ and ${\rm MS}(s,s')$, respectively, in this paper. 
Here, $s$ is a scaling factor that gives the number of sites on each side in the smallest unit. 
We then create empty sites, the number of which is $s'$ on each side also in the smallest unit (Fig.~\ref{SC(3,1)}). 
For the spectral dimension $d_{\rm s}$, our results are consistent with the earlier prediction for the upper and lower bounds within the margin of error~\cite{barlow1999brownian} (See Table~\ref{dimension and target sites}).

\section{\label{sec:quantum search algorithm}spatial search using flip-flop walk}

In this study, we use a spatial search algorithm with flip-flop walk~\cite{patel2012search,tamegai2018spatial}. 
A state $| \Psi \rangle$ obeys a discrete time evolution performed by an oracle operator $\hat R$ and a flip-flop walk operator $\hat W$, given by 
\begin{equation}
    \ket{\Psi(t)}=(\hat W\hat R)^{t}\ket{\Psi(t=0)}. 
    \label{time evolution operator}
\end{equation} 
The state  $\ket{\psi(t)}\equiv\sum_{{{\bf x},{\bf l}}}a_{{\bf x},{\bf l}} (t) \ket{{\bf x}}\otimes\ket{{\bf l}}$ is defined in the Hilbert space ${\mathcal H}_{\rm{search}}\equiv {\mathcal H}_{N}\otimes {\mathcal H}_{K}$, where $\ket{\bf x}\in {\mathcal H}_N$ is associated with the position degree of freedom, and $\ket{\bf l}\in {\mathcal H}_K$ is associated with the internal degree of freedom with the link vector ${\bf l}$ that orients a nearest neighbor site. 

The oracle operator is given by $\hat R=\hat R_N\otimes \hat I_K$ with $\hat R_N\equiv \hat I_N-2\ket{{\bf x}_0}\bra{{\bf x}_0}$, where $\hat I_N$ and $\hat I_K$ are the identity operators in ${\mathcal H}_N$ and ${\mathcal H}_K$, respectively, and ${\bf x}_0$ is the position of a target. 
The operator $\hat R$ plays a role of an attractive potential that enhances the finding probability at the marked vertex.  
The flip-flop walk operator $\hat W$ is composed of a Grover diffusion operator $\hat G$ and a flip-flop shift operator $\hat S$, i.e., $\hat W=\hat S\hat G$. 
The Grover diffusion operator \cite{grover1997quantum} evolves the probability amplitude as follows:
\begin{equation}
    a_{{\bf x},{\bf l}}\xrightarrow{\hat G}\frac{2}{k}\sum_{{\bf l}'}a_{{\bf x},{\bf l}'}-a_{{\bf x},{\bf l}}.
\end{equation}
We employ $k=4$ for a Sierpinski carpet and $k=6$ for a Menger sponge. 
The flip-flop shift operator $\hat S$ propagates a probability amplitude along its link direction with the link vector flipped:
\begin{equation}
\ket{{\bf x}}\otimes\ket{{\bf l}}\xrightarrow{\hat S}\ket{{\bf x}+{\bf l}}\otimes\ket{-{\bf l}}. \label{flip-flop walk} 
\end{equation}
When we consider a fractal lattice, there are missing sites (for example, the gray area in Fig.\ref{SC(3,1)}). In this case, we employ the following~\cite{tamegai2018spatial}: 
\begin{equation}
\ket{{\bf x}}\otimes\ket{{\bf l}}\xrightarrow{\hat S}\ket{{\bf x}}\otimes\ket{{\bf l}}.
\label{flip flop boundary}
\end{equation}
As an initial state, we use the uniform superposition state:
\begin{equation}
    \ket{\psi(t=0)}=
    \frac{1}{\sqrt{Nk}}\sum_{\bf x}^{}\sum_{\bf l}^{}\ket{{\bf x}}\otimes{\ket{{\bf l}}}.
\label{init psi}
\end{equation} 
The probability $P({\bf x},t)$ at a time step $t$ on a site ${\bf x}$ is given by
\begin{equation}
P({\bf x},t) = \sum_{\bf l}|\left(\bra{{\bf x}}\otimes\bra{\bf l}\right)\ket{\Psi(t)}|^{2}
=\sum_{\bf l}|a_{{\bf x},{\bf l}}(t)|^2.
\label{marked probability}
\end{equation}

In the numerical simulation, we follow the formalism of discrete-time quantum walk based on Eq.~(\ref{time evolution operator})--(\ref{marked probability}).
The time evolution of the probability amplitude $a_{{\bf x},{\bf l}} (t)$ is updated by following the rules of the oracle operator $\hat R$, and the flip-flop walk operator $\hat W$, where a set of operators $\hat W \hat R$ gives the discrete-time evolution $| \Psi (t+1) \rangle = \hat W \hat R | \Psi (t) \rangle$. The initial probability amplitude is prepared as $a_{{\bf x},{\bf l}} (t = 0) = 1/\sqrt{Nk}$, which corresponds to the uniform superposition state in Eq.~(\ref{init psi}).

\section{\label{sec:results}simulation results}

\begin{figure}[tbp]
          \includegraphics[width=80mm]{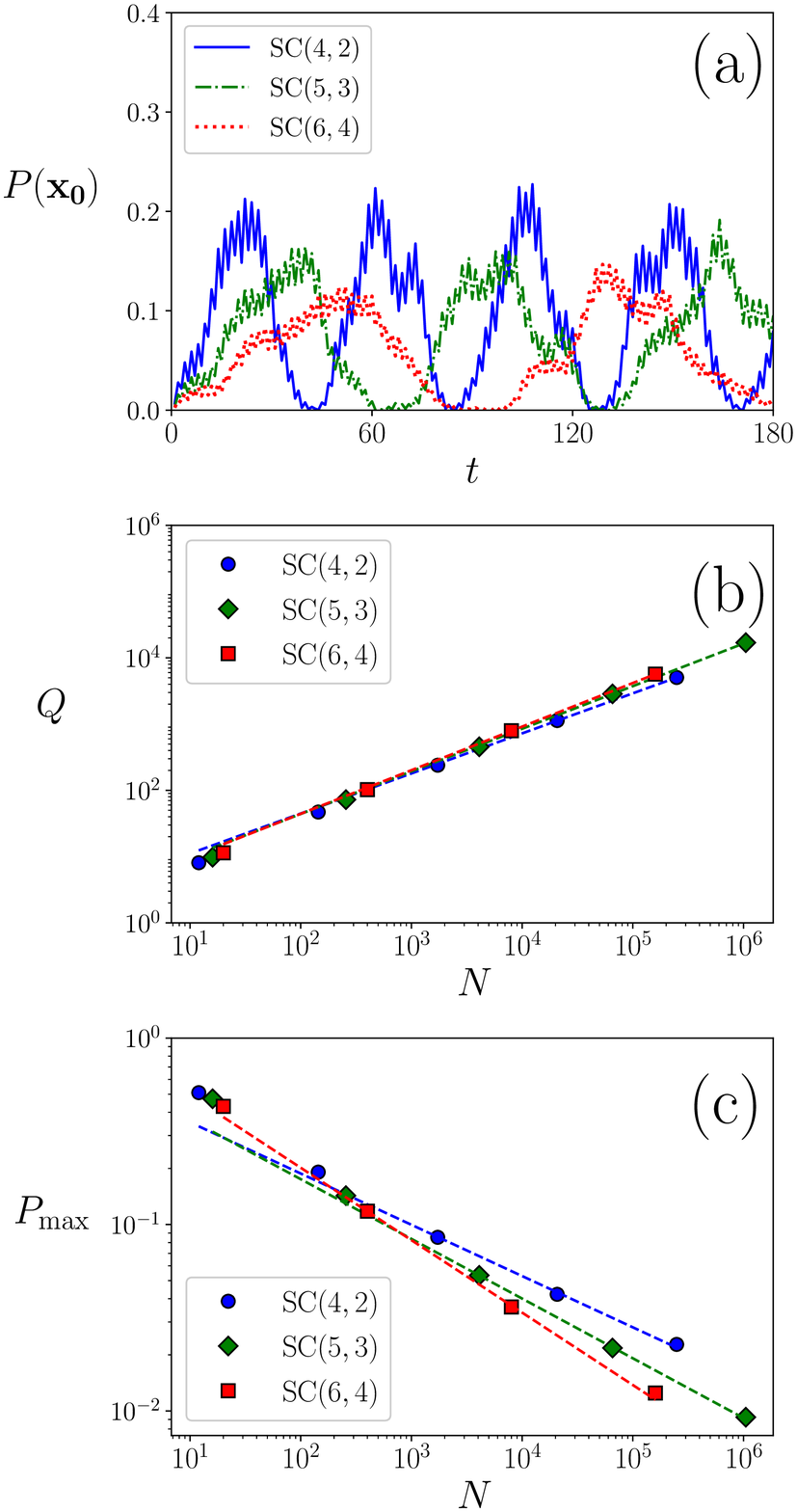}
    \caption{(Color online) Spatial search on extended Sierpinski carpets. 
    (a) Time-dependence of the probability at a marked site, $P({\bf x}_0,t)$, at the stage $S=2$. 
    (b) Site-number dependence of the optical oracle call $Q$, with a fitting $Q=O(N^\beta)$. 
    The fitting is performed in data from $S=1$ through $5$ for SC$(4,2)$, and from $S=1$ through $4$ for SC$(5,3)$ and SC$(6,4)$. 
    (c) Site-number dependence of the mean value of the maximum probability at a marked vertex $P_{\rm max}$, with a fitting $P_{\rm max}=O(N^{-\alpha})$. 
    The fitting is performed in data from $S=3$ through $5$ for SC$(4,2)$ and SC$(5,3)$, and from $S=2$ through $4$ for SC$(6,4)$.
    In this simulation, we employed the periodic boundary condition.
    }
    \label{fig:q_sc}
\end{figure}

\begin{figure}[tbp]
          \includegraphics[width=80mm]{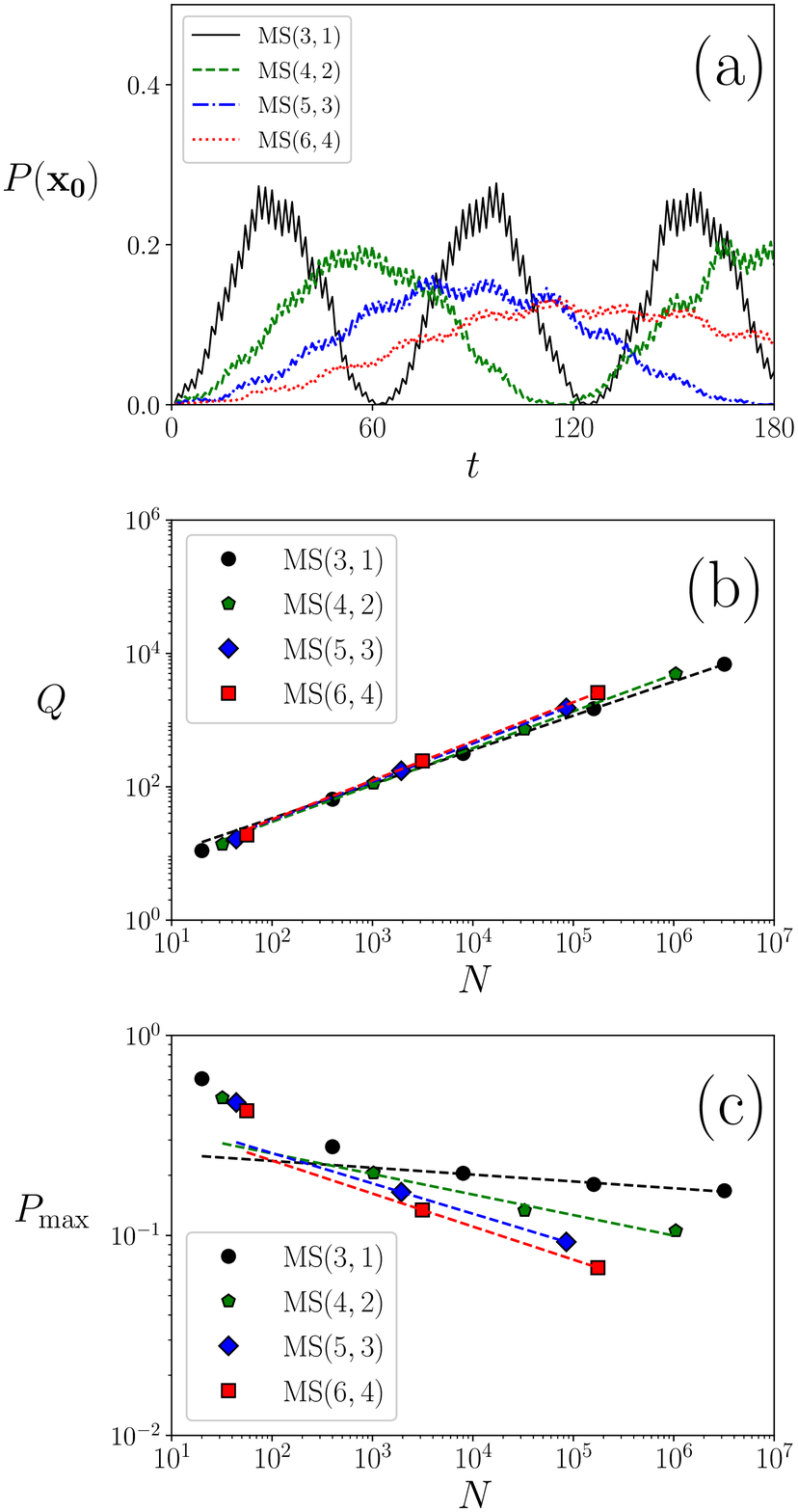}
              \caption{(Color online) Spatial search on conventional and extended Menger sponges. 
    (a) Time-dependence of the probability at a marked site, $P({\bf x}_0,t)$, at the stage $S=2$. 
    (b) Site-number dependence of the optical oracle call $Q$, with a fitting $Q=O(N^\beta)$. 
    The fitting is performed in data from $S=1$ through $5$ for MS$(3,1)$, from $S=1$ through $4$ for MS$(4,2)$, and from $S=1$ through $3$ for MS$(5,3)$ and MS$(6,4)$. 
    (c) Site-number dependence of the mean value of the maximum probability at a marked vertex $P_{\rm max}$, with a fitting $P_{\rm max}=O(N^{-\alpha})$. 
    The fitting is performed in data from $S=3$ through $5$ for MS$(3,1)$, from $S=2$ through $4$ for MS$(4,2)$, and from $S=2$ through $3$ for MS$(5,3)$ and MS$(6,4)$.
    In this simulation, we employed the periodic boundary condition.}
    \label{fig:q_ms}
\end{figure}

Figure~\ref{localized vertex} shows a typical example of the time evolution of the probability on the Sierpinski carpet SC$(3,1)$. 
The probability is concentrated at the target site and then diffuses, repeatedly (Sierpinski carpets in Fig.~\ref{fig:q_sc} (a) and Menger sponges in Fig.~\ref{fig:q_ms} (a)).

In order to analyze the scaling law, we perform numerical simulation about $100,000$-$1,000,000$ steps for Sierpinski carpets, and $5,000$-$100,000$ steps for Menger sponges. 
The optimal number of an oracle call $Q$, which is a characteristic period of the probability at a target site ${\bf x}_0$, is evaluated from a frequency giving the maximum intensity of the Fourier transformation of $P(t,{\bf x}_0)$. 
We also determine the mean value of the maximum probability $P_{\rm max}$. 
Here, a data set of $P(t,{\bf x}_0)$ is chronologically grouped with a period $Q$, and then the maximum values are evaluated in each group. By using this data set of maximum values, we calculate the mean value of the maximum probability $P_{\rm max}$. 
The optimal number of oracle call $Q$ and the mean value of maximum probability $P_{\rm max}$ are shown as functions of the number of sites $N$ in Figs.~\ref{fig:q_sc} (b) and (c) for Sierpinski carpets, and in Figs.~\ref{fig:q_ms} (b) and (c) for Menger sponges. 
We analyze the scaling law by supposing the relation $Q=O(N^\beta)$ and $P_{\rm max}=O(N^{-\alpha})$. The exponents $\alpha$ and $\beta$ obtained from the numerical data are summarized in Table~\ref{scaling exponent1}.

\begin{figure}[tbp]
    \centering
    \includegraphics[width=70mm]{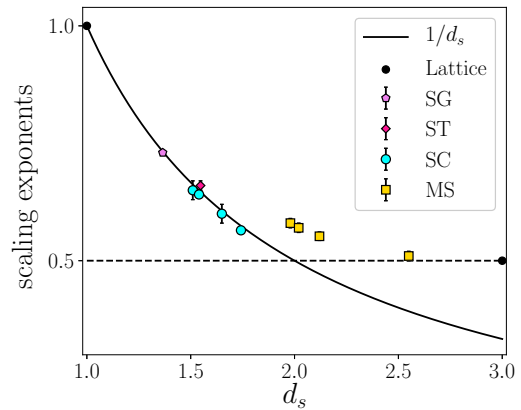}
    \caption{(Color online) Spectral dimension dependence of scaling exponents $\beta$, defined by $Q=O(N^\beta)$. This figure includes our numerical data obtained from Figs.~\ref{fig:q_sc}~(b) and~\ref{fig:q_ms}~(b) for the Sierpinski carpet (SC) and the Menger sponge (MS), respectively. It also includes the data of the Sierpinski carpet (SC), Sierpinski gasket (SG) and Sierpinski tetrahedron (ST) reported in Refs.~\cite{patel2012search,tamegai2018spatial} for non-integer dimensions as well as the data of hypercubic lattices (Lattice) in Refs.~\cite{childs2004spatial,aaronson2003quantum,Childs2004Dirac,ambainis2005coins,Patel2010I} for integer dimensions. The exponent follows the relation $\beta = 1/d_{\rm s}$ for $d_{\rm s} < 2$, 
and saturates at $\beta = 1/2$ for $d_{\rm s} > 2$. For the spectral dimension close to the critical dimension $d_{\rm s} \simeq 2$, the exponent $\beta$ does not follow the relation $\beta = 1/d_{\rm s} = 1/2$, which may suggest the necessity of the logarithmic correction.
}
    \label{b_ds}
\end{figure}

Patel and Raghunathan proposed the scaling hypothesis (\ref{scaling formula 2}), given in the form $Q \ge {\rm max} \{N^{1/d_{\rm s}},\pi\sqrt{N}/4\}$~\cite{patel2012search}. 
In order to check this conjecture, we plot the scaling exponent $\beta$, given by $Q \propto N^\beta$, as a function of the spectral dimension $d_{\rm s}$ in Fig.~$\ref{b_ds}$, which are obtained from data shown in Fig.~\ref{fig:q_sc}~(b) for the Sierpinski carpet and Fig.~\ref{fig:q_ms}~(b) for the Menger sponge, both of which give non-integer spectral dimensions. 
Figure~$\ref{b_ds}$ also includes the results of the hypercubic lattices~\cite{childs2004spatial,aaronson2003quantum,Childs2004Dirac,ambainis2005coins,Patel2010I} for integer dimensions $d_{\rm s} = 1$ and $3$ as well as the data of the Sierpinski carpet, Sierpinski gasket and Sierpinski tetrahedron in Refs.~\cite{patel2012search,tamegai2018spatial} for non-integer spectral dimensions. 
A similar plot can be found in Ref.~\cite{Boettcher2018_012320}, where the exponent $\beta$ is plotted as a function of $d_{\rm s}$ for the dual Sierpinski gasket and the Migdal--Kadanoff and Hanoi networks. 

We find that the scaling hypothesis (\ref{scaling formula 2}) holds well except for $d_{\rm s} \simeq 2$. 
The relation $\beta = 1/d_{\rm s}$ holds for $d_{\rm s} < 2$, 
and the exponent is saturated as $\beta = 1/2$ for $d_{\rm s} > 2$. 
For $d_{\rm s} \simeq 2$, the exponent $\beta$ deviates from the line obtained from the scaling hypothesis (\ref{scaling formula 2}), which strongly suggests the emergence of the logarithmic correction as well as the criticality of the two-dimensional system~\cite{aaronson2003quantum,ambainis2005coins,Childs2004Dirac,childs2004spatial,tulsi2008faster,Patel2010I}. 

To study the scaling law in more detail, we summarize values $\beta$ and $1/d_{\rm s}$ in Table~\ref{scaling exponent1} for Sierpinski carpets and Menger sponges.  
In the Sierpinski carpets, where $d_{\rm s} <  d_{\rm f} < d_{\rm E} = 2$, we find that the scaling hypothesis $\beta = {1/d_{\rm s}}$ holds well. 
In the Menger sponges, where $2 \lesssim d_{\rm s} < d_{\rm f}< d_{\rm E} = 3$, the relation $\beta \simeq {1/d_{\rm s}}$ does not hold. In this case, we find $\beta \simeq 0.5$, which reproduces $Q = {\mathcal O} (\sqrt{N})$ as in the scaling law (\ref{scaling formula 2}). 
In particular, in the case where the spectral dimension $d_{\rm s}$ is sufficiently greater than $2$, such as the case of MS$(3,1)$, the relation $\beta = 1/2$ is well reproduced, which is consistent with the hypercubic lattice case in $d>2$~\cite{aaronson2003quantum,Childs2004Dirac,ambainis2005coins,Patel2010I}. 
In the case where $d_{\rm s} \simeq 2$, the scaling hypothesis (\ref{scaling formula 2}) is not reproduced (See Fig.~$\ref{b_ds}$ and Table~\ref{scaling exponent1}). 
In $d_{\rm s} = 2$, the logarithmic correction is expected to emerge 
$Q \propto \sqrt{N} \ln^\epsilon N$~\cite{aaronson2003quantum,ambainis2005coins,Childs2004Dirac,childs2004spatial,tulsi2008faster,Patel2010I}. 
We analyzed the exponent of this logarithm close to $d_{\rm s} = 2$, and obtained the following fitting results: 
$\epsilon = 0.62(9)$ for MS$(4,2)$, 
$\epsilon = 0.689(5)$ for MS$(5,3)$, 
and 
$\epsilon = 0.84(3)$ for MS$(6,4)$.

\begin{table}[tbp]
    \centering
    \renewcommand{\arraystretch}{1.1}
    {\tabcolsep=1.8mm 
    \begin{tabular}{lrrrr}\hline\hline
& ${\beta}$ & ${1/d_{\rm s}}$ & ${\alpha}$ & ${\alpha-2\beta+1}$\\\hline
SC(3,1)\cite{tamegai2018spatial}& 0.5647(6)& 0.572(3)& 0.155(2)&0.026(2)\\
SC(4,2)&0.60(2)&0.606(7)&0.275(7)&0.04(2)\\  
SC(5,3)& 0.64(1)& 0.64(1)&0.320(3)&0.04(2)\\ 
SC(6,4)& 0.65(2)& 0.666(8)&0.38(1)&0.08(4)\\\hline

MS(3,1)& 0.51(1)& 0.392(9)&0.0350(4)&0.01(2)\\
MS(4,2)& 0.552(9)& 0.47(1)&0.10(4)&0.00(4)\\  
MS(5,3)& 0.57(1)& 0.492(9)&0.15(2)&0.01(2)\\
MS(6,4)& 0.58(1)& 0.50(2)&0.16\phantom{(x)}&0.00(2)\\\hline\hline
    \end{tabular}}
    \caption{Scaling exponents $\alpha$ and $\beta$ of  the scaling laws $P_{\rm max}=O(N^{-\alpha})$ and $Q=O(N^\beta)$ for the spatial search on Sierpinski carpets and Menger sponges studied in Fig.~\ref{fig:q_sc} and Fig.\ref{fig:q_ms}.  The numbers in parenthesis denote errors.  For the exponent $\alpha$ in MS$(6,4)$, the error cannot be evaluated, where fitting is performed only from data in two stages $S=2$ and $3$.}
    \label{scaling exponent1}
\end{table}

\begin{table}[tb]
    \centering
    \renewcommand{\arraystretch}{1.2}
    {\tabcolsep=1.8mm \scriptsize{
    \begin{tabular}{lrrrrr}\hline\hline
 &$\gamma$& ${\gamma'}$& ${\gamma''}$& ${\gamma-\gamma'}$& ${\gamma-\gamma''}$\\\hline
SC(3,1)\cite{tamegai2018spatial}&0.642(1)& 0.645(8)&0.317(4)&0.007(8)&0.325(4)\\
SC(4,2)&0.73(2)& $-$0.56(2)&0.72(1)&1.30(2)&0.01(2)\\  
SC(5,3)&0.80(1)& 1.74(4)&0.79(2)&2.54(4)&0.01(2)\\ 
SC(6,4)&0.84(2)& $-2.83(2)$&0.84(1)&3.67(2)&0.00(2)\\\hline

MS(3,1)&0.52(1)&1.00(3)&0.50(2)&${-0.48(3)}$&0.02(1)\\
MS(4,2)&0.60(1)&$-$0.44(3)&0.78(1)&1.04(3)&${-0.18(2)}$\\  
MS(5,3)&0.64(1)&$-$1.64(2)&0.85(1)&2.28(2)&${-0.20(2)}$\\
MS(6,4)&0.66(1)&$-$2.76(4)&0.87(2)&3.42(4)&${-0.20(2)}$\\\hline\hline
    \end{tabular}}}
    \caption{Scaling of the effective number of oracle calls $Q/\sqrt{P_{\rm max}} = {\mathcal O}(N^\gamma)$. The exponents $\gamma'$ and $\gamma''$ are given in Eqs. (\ref{c'}) and (\ref{C''}), respectively. }
    \label{scaling exponent2}
\end{table}

Patel and Raghunathan proposed another scaling relation~\cite{patel2012search}
\begin{equation}
\alpha=2\beta-1, 
    \label{scaling_a2b1}
\end{equation}
in the context of the Tulsi's amplification algorithm for the success probability. 
In Table~\ref{scaling exponent1}, we find the relation $\alpha-2\beta+1\simeq 0$ seems to hold. 
In particular, we find that the scaling relation (\ref{scaling_a2b1}) holds within the margin of error for the Menger sponges with $2< d_{\rm f} < d_{\rm E}=3$.

Finally, we investigate the scaling of the effective number of oracle calls $Q/\sqrt{P_{\rm max}}={\mathcal O}(N^\gamma)$, where $\gamma$ is given by 
\begin{equation}
\gamma = \beta + \frac{\alpha}{2}. 
 \label{c}
\end{equation} 
In the conventional Sierpinski carpet SC$(3,1)$, the Sierpinski gasket, and the Sierpinski tetrahedron, which corresponds to the case where $s'=1$, the scaling hypothesis is given by 
\begin{equation}
 \gamma = \frac{d_{\rm s}}{d_{\rm E}-1} + d_{\rm f} - {\rm s},
 \label{c'}
\end{equation} 
which holds within an error as shown in Ref.~\cite{tamegai2018spatial}. 
In the extended Sierpinski carpet SC$(2+s',s')$ for $s'>1$, 
we find that the hypothesis (\ref{c'}) does not hold (Table.~\ref{scaling exponent2}), 
which suggests that we may need another scaling law. 
Here, we propose a hypothesis for $s' > 1$ analogous to (\ref{c'}), given in the form 
\begin{equation}
\gamma = \frac{1}{2}\left(\frac{d_s}{d_E-1}+d_f-\frac{s}{s'}\right). 
\label{C''}
\end{equation}

As shown in Table~\ref{scaling exponent2}, the scaling hypothesis (\ref{C''}) is reproduced within an error for $s' > 1$ in the extended Sierpinski carpets, where $d_{\rm s} < d_{\rm f} < d_{\rm E} = 2$. 
In the conventional Menger sponge MS$(3,1)$ for $d_{\rm s} = 2.55(6)$, 
we find that the relation (\ref{C''}) approximately holds. 
However, in the case of the extended Menger sponges MS$(2+s',s')$ with $s'>1$ for $d_{\rm s} \simeq 2$, we find that both scaling conjectures (\ref{c'}) and (\ref{C''}) do not hold. Indeed, close to $d_{\rm s} =2$, the system may be in the critical regime.

In short, there may be the scaling law for the effective number of oracle calls in the Sierpinski carpet for $d_{\rm s} < 2$, which is given by quantities characterizing a fractal geometry, such as the Euclidean dimension, the fractal dimension, the spectral dimension and the scaling factor. 
For the conventional Sierpinski carpet with $s'=1$, 
the scaling relation (\ref{c'}) holds reasonably well. 
For the extended Sierpinski carpet with $s'>1$, 
the scaling relation (\ref{C''}) holds well within the margin of error. 
For the Menger sponges, on the other hand, the scaling relation (\ref{scaling_a2b1}) holds well within the margin of error. 

An important issue for future study is to prove the conjectures \eqref{scaling formula 2}, \eqref{c'}, and \eqref{C''}. 
One of an interesting approach may be to employ idea of the renormalization~\cite{Boettcher2018_012309,Boettcher2018_012320}. 
Since the relations \eqref{c'} and \eqref{C''} are reproduced very well, 
we naturally expect that there should be a beautiful mathematical structure behind the quantum spatial search on a fractal geometry. 
We further need to explore the scaling law for a non-integer dimension close to $d_{\rm s} = 2$, where the logarithmic correction may be expected.  Finally, the reason of the appearance of the factor $1/2$ in \eqref{C''}, compared with \eqref{c'}, must be clarified. One of the possible origin of the factor $1/2$ might be given by $1/(s-s')$ for $s' > 1$, although we have studied only the case for $s-s' = 2$ in this paper.

In recent experiments, the quantum walk can be performed in an optical lattice~\cite{Karski:2009jc}, an ion trap~\cite{Zahringer:2010bs}, and photonic sytems~\cite{Xiao2018it,Xiao2017iw,Schreiber2010cl,Schreiber2012wj}. 
Furthermore, single electron in real space is accessible in Sierpinski gasket fabricated on Cu$(111)$ by using scanning tunneling microscopy and spectroscopy~\cite{kempkes2019design}. 
With these developments, we hope that the quantum spatial search on the fractal lattice will become accessible in near future to confirm the scaling behavior of the finding probability at a specific target. 
Indeed, the spatial search may be emulated by the continuous-time Schr\"odinger dynamics on the lattice with a single attractive potential site~\cite{Grover2001,Agliari2010dw}. 
For example, suppose that an optical lattice with a fractal geometry is prepared. 
Then, trap a single atom in it for making the uniform superposition state in the fractal lattice. 
It may be more preferable to use a macroscopic matter wave of the Bose--Einstein condensate (BEC). 
By analyzing the time-evolution of the finding probability or the density profile of the BEC with use of the single atom addressing technique~\cite{Bakr:vp,Weitenberg:vm}, the scaling behavior depending on the topological structure of the lattice may be studied as in the present paper. 
Dynamical properties at a specific local site may be governed by the global structure of the fractal lattice, such as the dimensions, and the scaling factor. 
We hope that the present study further promotes interdisciplinary study of complex networks, mathematics and quantum mechanics on the fractal geometry.

\section{\label{sec:conclusion}conclusions}

Many real-world systems, such as the World-Wide-Web, social and biological networks, exhibits the complex network structure. 
Developing an efficient search algorithm on such complex networks is an important issue, and one of the efficient ways is to apply the spatial search by quantum walk. 
An interesting point is that these complex networks may have the self-similarity in spite of the ``small-world" property. 
In addition to the important practical problem of appying quantum spatial search to each real-world system composed of complex networks, there naturally arises the fundamental question, namely, what kind of scaling law is behind the quantum spatial search on self-similar networks, such as fractal lattices. 

We investigated the spatial search using quantum walk on various kinds of fractal lattices, such as Sierpinski carpets and Menger sponges. 
We first studied the conjecture by Patel and Raghunathan proposed for the Sierpinski gasket and tetrahedron, and corroborated their conjecture for other fractal lattices, where the exponent for the optimal number of oracle calls with respect to the number of sites is given by the inverse of the spectral dimension $d_{\rm s}$ for $d_{\rm s}<2$, and approaches $1/2$ for $d_{\rm s}$ being far from $2$. For $d_{\rm s} \simeq 2$, the deviation from their conjecture emerges, which supports the fact that the two-dimensional system is critical and the logarithmic correction may be needed. 
We also studied the relation between the exponent for the optimal number of oracle calls and the exponent for the mean value of the maximum probability at a specific target.  

Finally, we proposed the scaling conjecture of the effective number of oracle calls, which is given by characteristic quantities of the fractal geometry, such as the Euclidean dimension, fractal dimension, spectral dimension, as well as the scaling factor. The relation for the extended Sierpinski carpets proposed in this paper is similar but slightly different from the relation propoesd by Tamegai and two of the authors for the Sierpinski gasket, Sierpinski tetrahedron, and conventional Sierpinski carpet. 
Nevertheless, it may be indeed a fact that the quantum spatial search on the fractal lattice is surprisingly subject to the combination of the characteristic quantities of the fractal geometry. 
It is an open question why the scaling law of the effective number of oracle calls is given by such combination of the dimensions and the scaling factor. 
Mathematical proofs of these scaling relations have not been also provided yet, and are left for a future study.

\begin{acknowledgments}
S.W. was supported by JSPS KAKENHI Grant No. JP16K17774.  T.N. was supported by JSPS KAKENHI Grant No. JP16K05504. 
We thank Apoorva Patel for useful discussion about the scaling hypothesis about the number of oracle calls. 
We also thank T. Akimoto for discussion about the classical random walk, and the fractal lattice. 
\end{acknowledgments}

\bibliography{mybib}

\end{document}